\documentclass[letter]{IEEEtran}
\IEEEoverridecommandlockouts
\usepackage{amsmath,amssymb,epsfig,authblk,cite,booktabs}
\usepackage{color,hyperref,graphicx,cite}
\usepackage{ifpdf}
\ifpdf
  \DeclareGraphicsExtensions{.pdf,.png,.jpg}
\else
  \DeclareGraphicsExtensions{.eps}
\fi
\usepackage{epstopdf}

\hypersetup{
    colorlinks,%
    citecolor=blue,%
    filecolor=blue,%
    linkcolor=blue,%
    urlcolor=blue
}
\definecolor{bluecolor}{rgb}{0,0.,1.}

\definecolor{redcolor}{rgb}{.7,0.,0.}

\newcommand{\vect}[1]{\boldsymbol{#1}}

\usepackage[normalem]{ulem}

\newcommand{\es}[1]{\begin{equation}\begin{split}#1\end{split}\end{equation}}

\newcommand{\argmax}{\operatornamewithlimits{argmax}}

\newcommand{\squeezeup}{\vspace{-1.3mm}}
\IEEEoverridecommandlockouts

\makeatletter
\def\bstctlcite{\@ifnextchar[{\@bstctlcite}{\@bstctlcite[@auxout]}}
\def\@bstctlcite[#1]#2{\@bsphack
  \@for\@citeb:=#2\do{%
    \edef\@citeb{\expandafter\@firstofone\@citeb}%
    \if@filesw\immediate\write\csname #1\endcsname{\string\citation{\@citeb}}\fi}%
  \@esphack}
\makeatother

\begin{document}
\title{\huge{Reconfigurable Intelligent Surfaces and Machine Learning for Wireless Fingerprinting Localization}}

\author{Cam Ly Nguyen, Orestis Georgiou, ~\IEEEmembership{Member,~IEEE}, and Gabriele Gradoni,~\IEEEmembership{Member,~IEEE}

\thanks{C.L. Nguyen is with the Wireless System Laboratory, Corporate Research \& Development Center, Toshiba Corporation, Kawasaki, Japan. 
O. Georgiou is with the Department of Electrical and  Computer  Engineering, University of Cyprus, Nicosia, Cyprus.
G. Gradoni is with the School of Mathematical Sciences and the Department of Electrical and Electronic Engineering, University of Nottingham, University Park, United Kingdom.}
}

\maketitle

\bstctlcite{IEEEexample:BSTcontrol}

\begin{abstract}
Reconfigurable Intelligent Surfaces (RISs) promise improved, secure and more efficient wireless communications.
We propose and demonstrate how to exploit the diversity offered by RISs to generate and select easily differentiable radio maps for use in wireless fingerprinting localization applications.
Further, we apply machine learning feature selection methods to prune the large state space of the RIS, thus reducing complexity and enhancing localization accuracy and position acquisition time.
We evaluate our proposed approach by generation of radio maps with a novel radio propagation modelling and simulations.
\end{abstract}


%%%%%%%%%%%%%%%%%%%%%
\squeezeup
\section{Introduction} 
\label{sec:introduction}
%RIS introduction

The reconfigurable intelligent surface (RIS) is a promising  technology for enhancing connectivity and beyond in the sixth generation of mobile networks (6G) \cite{basar2019wireless}. 
The propagation of waves through RIS-empowered environments becomes agile in the sense that it can be controlled to perform energy focusing with wavelength scale accuracy. 
RIS devices are compact, quasi-passive electronics mirrors that, upon illumination by a primary electromagnetic (EM) signal source, e.g., an access point (AP), operate to create tailored wave back-scattering to optimise specific signal characteristics at a dynamic target, e.g., a mobile user (MU) terminal. 
The back-scattering is realised through an array of $N$ metallic elements, e.g., dipoles or patches, which are loaded with tuneable circuitry. 
The flexibility of the RIS, united with their easy and massive deployment has potentials to create an unprecedented capillary network for distributed sensing and computation \cite{di2019smart}. 
However, this stands on the capability of localizing and positioning target MU devices accurately within the propagation environment.

Improved localization and positioning of MUs and other internet of things (IoT) devices using RIS capabilities is a key enabling element of 6G systems.
It is designed to coexists with ubiquitous communications and to support a number of novel applications.
These include: imaging for biomedical and security applications; 
applications of simultaneous localization and mapping (SLAM) to automatically construct maps of complex indoor environments; 
passive sensing of people and objects; 
using location information as a big data source, guiding and predicting the human-digital ecosystem; 
the coexistence and cooperation between sensing, localization and
communication, leading to one device with a three-fold functionality; 
and finally, the use of location information to boost security and trust in 6G connectivity solutions \cite{bourdoux20206g}.

Our aim here is to provide novel methods and models for RIS-enabled wireless localization that leverage machine learning approaches to support the aforementioned 6G applications and use cases.
Moreover, unlike with most localization and positioning algorithms which are based on the path loss model, we use here a unified end-end communication model based on impedance matrices of thin wire antennas \cite{Gradoni2020} to calculate the composite AP and RIS induced EM field and build fingerprint maps for indoor localization \cite{zafari2019survey}.
This is important because conventional methods for indoor localization are unreliable in RIS-enabled environments. For instance in time-of-arrival (ToA) and in angle-of-arrival (AoA) based techniques it is difficult for the MU to determine whether a signal came from the AP or the RIS \cite{mu2020intelligent}. 
Furthermore, in received signal strength information (RSSI) based techniques, the conventional propagation models, e.g., Friis model, need to account for the ability of the RIS to completely change (beamform, scatter, null) the radio map spatial power distribution.

\begin{figure}[t]
\centering
\includegraphics[ width=0.6\linewidth]{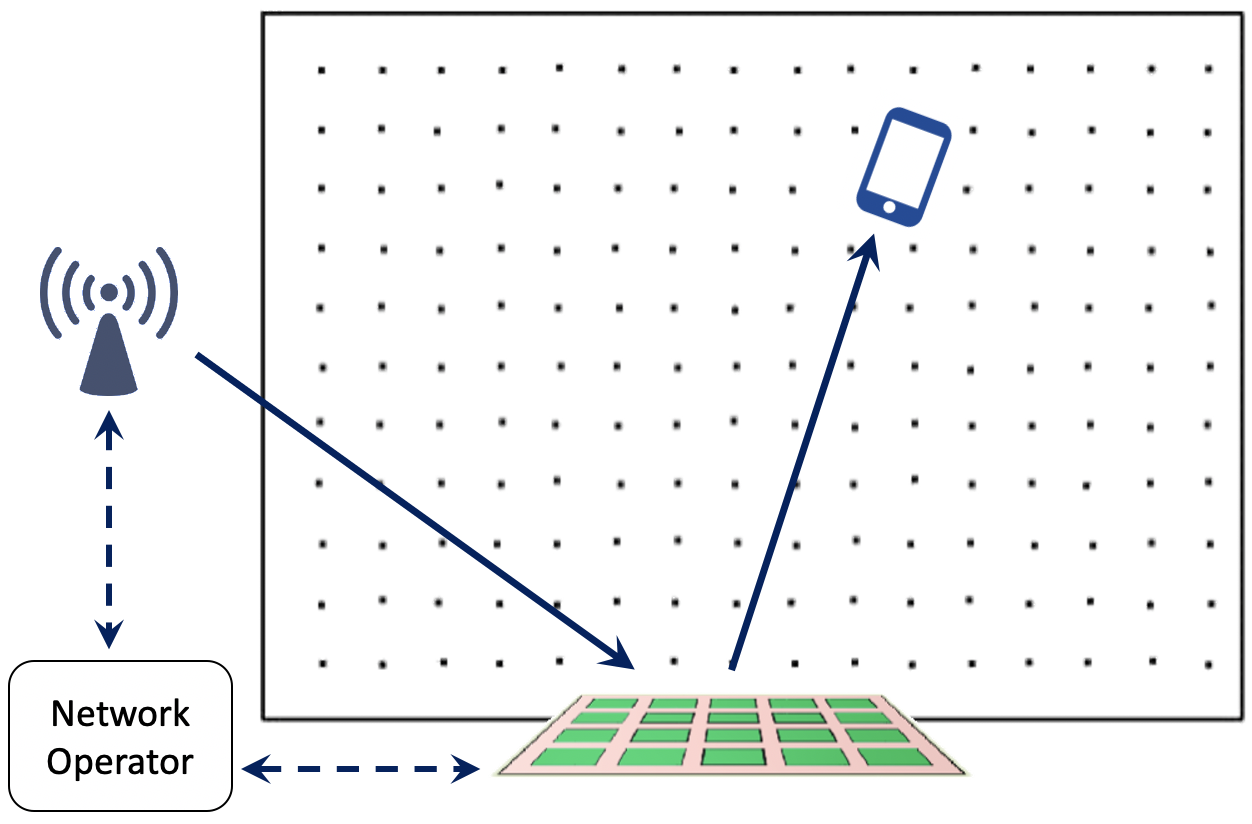}
\caption{System model for RIS-assisted localization. It consists of a transmitter and a RIS which are connected to a network operator.
A MU attempts to self-localize relative to a radio map sampled along a grid of points during an offline measurement campaign.} 
\label{fig:system}
\end{figure}

Our main contribution is the use of machine learning feature selection methods to a wireless fingerprinting localization problem that is further enabled by a RIS.
We exploit the degrees of freedom injected by the RIS to train our model and enhance localization accuracy performance. 
We validate our proposed method using novel radio propagation modelling simulations \cite{Gradoni2020} which are appropriate for RIS-enabled settings.

\begin{figure*}[t]
\centering
\includegraphics[ width=0.9\linewidth]{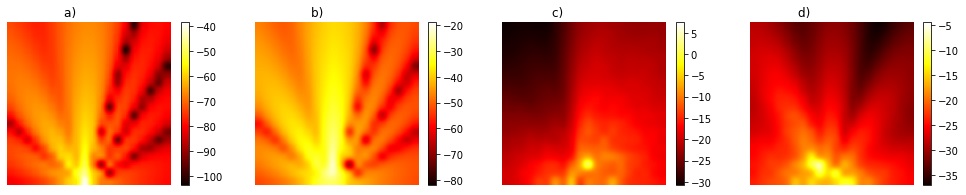}
\caption{Example of four Radio Maps in noise-free environments under difference configurations of RIS with the AP and the RIS situated as in Figure \ref{fig:system}. 
a) and b) apply two different uniformly increasing impedance values to the RIS elements thus creating different beamsteering radio maps, c) applies a random impedances chosen from a uniform distribution, and d) applies a constant impedance thus simply reflecting the incoming radiation from the AP.} 
\label{fig:map}
\end{figure*}

%%%
\squeezeup
\section{Background and Related Works}
\label{sec:related}

%1) a few works on RIS
RIS technologies have given rise to the concept of ``smart radio environments" thus unlocking the engineering of the wireless propagation environment itself \cite{di2019smart}.
In doing so, already many works have studied how to alleviate shortcomings due to multi-path fading or blockages in non line-of-sight (NLOS) settings \cite{bjornson2019intelligent}, whilst others have reported boosted multi-user downlink rates \cite{di2020hybrid} and MIMO diversity as well as throughput gains while remaining energy efficient  \cite{huang2020holographic}. 
While there has been much interest in the connectivity and coverage benefits afforded by RIS-enabled communications, very little work has been done with respect to wireless localization.

%localization in 5G
Wireless localization and positioning is a key element in cellular and WiFi networks, enabling a number of location based services for commerce and security, supporting autonomous vehicles, and providing context awareness both indoors and outdoors \cite{zafari2019survey}.
RSSI multilateration techniques map signal strength to distance to some known anchor set of APs using a path loss model, and are thus simple and cost-efficient to deploy but suffer from poor accuracy due to fading and attenuation uncertainties. 
%Additional channel configuration information (CSI) and ultra-wideband (UWB) antennas can somewhat mitigate multipath and indoor noise.
AoA based approaches leverage multiple antennae arrays to resolve the relative angles of incoming signals from APs and use triangulation to self-localise.
AoA approaches require additional signal processing at the MU while localization accuracy deteriorates at larger distances or in multipath environments even if the angular error is relatively small. 
ToA signal processing techniques require tight synchronisation, bandwidth, and high sampling rates while still suffering from multipath induced errors.
%we could reduce some of the background info if space saving is needed

Fingerprinting techniques mitigate such issues by first building a dense spatial database of RSSI measurements during an offline phase (usually along a grid of $L$ locations), and then use real-time measurements obtained during the online phase to compare with the offline measurements and estimate the MU location.
One immediate overhead issue with fingerprinting approaches is that new radio maps are required every time the environment changes, but can be overcome with crowdsourcing techniques \cite{wu2014smartphones}.
A second issue is that in overly dense fingerprint radio maps, the RSSI difference between two neighboring sample points becomes smaller than the typical signal variations thus making them indistinguishable and the estimation of the correct position impossible.
Many authors have tried to addressed this shortcoming in fingerprinting localization through machine learning methods such as deep neural networks (DNN) \cite{nowicki2017low} and $k$-nearest neighbour (kNN) \cite{fang2018optimal} to improve classification accuracy, and increase robustness against noise variability and multipath effects. 
However,` the best way to improve accuracy is to simply add more APs.

RISs promise to inject new capabilities and unlock centimeter scale accuracy in wireless localization and positioning.
Unfortunately, work in this direction is still underdeveloped.
Hu et al. \cite{hu2017cramer} calculated lower bounds for point to point positioning through RISs. 
Huang et al. \cite{huang2019indoor} described a DNN-based method for online wireless configuration of the RIS based on fingerprint localization estimates that beam-steer onto the MU thus optimising its RSS.
He et al. \cite{he2020large} studied the theoretical performance limits of a single anchor MIMO system using a path loss LOS model while also evaluating the impact of the number of RIS elements, and proposed adaptive phase shifter designs based on hierarchical codebooks \cite{he2020adaptive}.
Finally, Zhang et al. \cite{zhang2020metaradar} proposed a method that modifies the fingerprint radio map and improves localization accuracy by making RSS values at adjacent data set locations have significant differences.
The method selects the best RIS configurations, which is the key idea to enhance localization accuracy, by assuming that real-time radio maps are built through mathematical path-loss models. 
%However, in practice, since signals suffer multipath shadowing, fading etc, mathematical pathloss models do not express accurate radio maps. 
However, this is difficult to be deployed in practice unless accurate real-time radio maps can be constructed in accurate simulation models, e.g., ray tracing. 

Here, we propose a practical fingerprinting localization method that can enhance the localization accuracy and time by optimizing the RIS configurations during the offline training phase, i.e., using real measurement data: 
the method first collects RSSI values corresponding to many different RIS configurations at every reference location. 
Then, a feature selection technique combined with a supervised learning technique is used to select the best subset of RIS configurations for accurate localization.
The newly reduced in size RIS configuration subset is then used in the online phase, thus offering improved accuracy and location acquisition time.

%Taking inspiration from \cite{zhang2020metaradar}, we propose that radio map differentiability is a key control parameter in RIS-enabled wireless localization and positioning, and propose a simple machine learning method for identifying and maximising this differentiability, and thus improve localization accuracy.
%we could remove or rephrase the contribution by Zhang.

%Wireless Localization
%Among RSSI-based localization methodologies, fingerprinting techniques \cite{} enhance accuracy. First, during the training phase, RSSI values, called \emph{fingerprints}, are collected at multiple points of interest (e.g. grid points over the space) to construct a RSSI database. 
%Each point in the map comprises multiple entries corresponding to signal strength to transmitters at different positions. 
%The observation at each point is the RSSI vector consisting of the strength of signals sent by multiple transmitters.
%Then, supervised learning method is used to estimate the receiver's position. 
%In these approaches, the receiver performs localization using RSSI values from multiple transmitters.   

%Our approach

%Since conventional RSSI-based approaches base on an unique propagation model where the signal strength %recorded with an omnidirectional antenna is related to the propagation distance, they require multiple transmitters. 

\squeezeup
\section{System Model and Problem Statement}
\label{sec:model}
%\section{RIS-Assisted Fingerprinting Localization}

In typical fingerprint localization approaches
%radio environments 
with multiple APs, each AP contributes towards one fingerprint. 
Thus, increasing the number of APs  generates a longer fingerprint vector that generally improves the localization accuracy via radio map differentiability and robustness \cite{yiu2017wireless}.
In a RIS-assisted environment however, using just one AP and the RIS, multiple fingerprints can be created through configuring the RIS in different ways. 
Figure \ref{fig:map} illustrates four example radio maps corresponding to four different configurations of the RIS in an indoor space shown in Figure \ref{fig:system}. 
Thus, by changing the configuration of the RIS, i.e., the load impedance of the dipoles, one can get a much more diverse set of radio maps.
To that end, this paper proposes a fingerprinting approach and machine learning (ML) combining with feature selection (FS) method to accurately estimate the MU localization coordinates using just one AP and one RIS. 

%We notice that by changing the configuration of the RIS (i.e. phases of the dipoles) we can get diverse radio signal maps? 
%[GG: Shall we say that map differentiability is the key ingredient here? ]

We consider a transmitter (e.g., an AP), a RIS, and a receiver (e.g., a MU),
%The AP has a line of sight (LOS) communication with the RIS, and the RIS has a LOS with the MU.
%For simplicity we shall assume that there is no LoS between AP and MU (see Figure \ref{fig:system}). 
and assume that the RIS and the AP are connected to a network operator that can also control the RIS configuration.
Each RIS configuration will generate a 
%slightly 
different radio map (see Figure \ref{fig:map}).
The RIS is usually made up of $N$ quasi-passive tuneable elements, 
often modelled as cylindrical thin wires of perfectly conducting patches, arranged periodically across a grounded dielectric substrate.
Due to hardware limitations, the complex values (changing amplitude and shifting phases of the reflected waves) that the $N$ load impedances of the RIS reflecting elements can assume, 
need to be quantized into $D$ discrete values.
Thus, the RIS can be electronically controlled into any one of $S= D^N$
%$S=(N\times D)!$ 
possible configurations. 
Note that $S$ is usually a very large number. 

Our model is intentionally simple but generic enough and can be expanded to multi-AP and multi-RIS in future studies.

When a localization request is %broadcasted 
sent by the MU, the network operator instructs the AP to transmit a sequenced burst of $M\leq S$ wireless messages periodically every $t$ millisecond (e.g., every 100 millisecond). 
%Note that $M$ is a pre-defined positive integer number, and it should not be too big. 
%This is because sending a large number of signals can cause: a delay in real-time localization (e.g. the receiver moves far away from the original position at which the location estimation phase starts); high communication cost.
Meanwhile, the network operator also configures the RIS by setting the load impedances periodically every $t$ milliseconds, such that the $m$-th transmitted message corresponds to the $m$-th configuration of the RIS ($1 \leq m \leq M$). % in terms of an array of complex-valued impedance loading the RIS unit cells. 
The MU located at some unknown position $x$ receives the $M$ messages from the transmitter and calculates their RSSI values to form an RSSI vector $\vect{R}_x= [R_1, R_2, ...., R_M]$. 
To reduce communication overhead, the MU then compares $\vect{R}_x$ to a database of radio maps created during the offline fingerprint measurement campaign and communicated to the MU by the AP, possibly with the assistance of the RIS.

There are a number of established algorithms available that compare and match the offline and online measurements, e.g., probabilistic, neural networks, nearest neighbours, etc. \cite{yiu2017wireless}.
To illustrate their basic principle one can consider the offline radio maps which are are essentially a set of $L$ RSSI vectors $\hat{\vect{R}}_l=\{\hat{R}_1, \hat{R}_2,\ldots, \hat{R}_M \}$ with $l\in[1,L]$ corresponding to one of the $L$ sampled location coordinates (usually along a grid) obtained during an offline measurement campaign, and compare them to a measured RSSI vector $\vect{R}_x$ through a permuted Pearson's correlation coefficient
\es{
l^*=
 \argmax_{l\in[1,L]} \Big( \max_{k\in[1,M]} \frac{\text{cov}(\vect{R}_x,\vect{\pi}^k \hat{\vect{R}_l})}{\sigma_{\vect{R}_x}\sigma_{\vect{\pi}^k\hat{\vect{R}}_l}} \Big)
\label{compare}} 
%\Blue{Ly: Actually, the kNN (k= 1) is much simpler than \eqref{compare}  $l^*= \argmax_{l\in[1,L]} ||\vect{R}_x- \vect{R}_l || $. We assume that in the localization phase, the MU receives consecutive M messages corresponding to M RIS configurations. Namely, $R_i$ corresponds to the $i$-th RIS configuration, meanwhile $\hat{R_i}$ also correspond to the $i$-th RIS configuration. Therefore, the RSSI distance is simply the Euclidean distance of the two RSSI vector.   Permuted Pearson correlation can be used under the assumption that the MU does not know the starting point of the RSSI vector, e.g. the AP and RIS do not synchronize with ecah other}.
where $\vect{\pi}^k$ is an $M\times M$ permutation matrix that cycles the elements of $\hat{\vect{R}}_l$ by $k$ positions to the left.
Thus, in \eqref{compare}, the inner max operation finds the largest Pearson's correlation coefficient when comparing the online measured RSSI vector $\vect{R}_x$ to a permutation of the $l_{\text{th}}$ offline radio map RSSI vector $\vect{\pi}^k \hat{\vect{R}_l}$, while the outer argmax operation returns the location $l^*$ with the most similar RSSI vector, thus identifying the most likely location of the MU.
Note that the Pearson's correlation coefficient is equivalent to the cosine similarity metric of centred (zero mean) vectors.
The $k$ nearest neighbour ($k$-NN) algorithm takes an average of the $k$ most similar locations.

Inspecting equation \eqref{compare} we observe that its computational complexity grows with $L$ and $M$.
Importantly, the process of acquiring (and maintaining) the radio map database is a labor intensive and time consuming effort which grows with $L$.
Also, the time needed for the AP and RIS to transmit and reconfigure the EM spatial distribution grows like $M\times t$.

Our aim is therefore to propose and evaluate methods for efficiently selecting the  $M\leq S$ \textit{best} RIS configurations while also reducing $L$ but not compromising localisation error.
In the next three sections we will describe how we generate realistic radio maps, in the absence of real measurement data, and propose two approaches for selecting the optimal RIS configuration subset, a heuristic state selection (HSS) and a machine learning feature selections (ML-FS) one.

\squeezeup
\section{Radio Map Generation \label{sec:radiomap}}

The RIS model implicitly account for the mutual coupling between elements, which is important for obtaining different path loss responses at every receiver location that are uniquely associated to an amplitude-phase mask of the RIS array.
Inherently, rather than the common phase-based unit cell control, this physics based model can describe the change in both the amplitude and phase of the locally impinging EM waves through 
the complex-valued loading impedances. 
Here, the unit cell of the RIS is a dipole, for which self- and mutual-impedance expressions are available in closed-form.

Considering the single-input single-output system in Figure \ref{fig:system}, we let $V_\text{AP}$ be the (complex-valued) voltage at the terminals of the AP signal generator, and $V_x$ the voltage at the terminals of the MU. 
Combining antenna theory with circuit network theory \cite{Gradoni2020}, an end-to-end channel model is obtained as 
$ V_x = \mathcal{H}_{\text{E2E}} \, V_\text{AP}$, where $\mathcal{H}_\text{E2E} = \mathcal{H}_\text{LOS} + \mathcal{H}_\text{VLOS}$, 
with virtual line-of-sight (VLOS) indicating the propagation path from AP to MU via the RIS. 
In the far-field of the RIS, i.e., for distances between AP or MU and RIS larger than a wavelength, Corollary 1 in \cite{Gradoni2020} yields $\mathcal{H}_\text{LOS} \approx Z_\text{RT}$ 
and  
\begin{equation}
    \mathcal{H}_\text{VLOS} \approx 
    \sum_{u=1}^N \, \sum_{v=1}^N \, 
    \mathcal{Z}_\text{RS,u} \, 
    \Phi_{uv} \, 
    \mathcal{Z}_\text{TS,v} 
\end{equation}
where $\Phi_{uv}$ is the reflection coefficient of the RIS element $(u,v)$, $\mathcal{Z}_\text{RS,u}$ the $u$-th element of the transfer impedance vector between RIS and MU, $\mathcal{Z}_\text{ST,v}$ the $v$-th element of the transfer impedance vector between RIS and AP, and $\mathcal{Z}_\text{RT}$ is the transfer impedance between AP and the MU.
Detailed expressions for the elements of the vectors $\mathcal{Z}_\text{RT}$, 
$\mathcal{Z}_\text{RS}$, $\mathcal{Z}_\text{ST}$, 
as well as the matrix $\Phi$ can be found in \cite{Gradoni2020}.
%Thus, 
%\Red{since we do not have experimental data at hand, - maybe we can skip this?} 
We can thus calculate the radio map at every location $l$ in the spatial grid used for fingerprinting through the received power $P_l = \left | \mathcal{H}_\text{E2E} \right |^2 \, P_\text{AP}$ plus some noise $X \sim \mathcal{N}(0, \sigma^2)$ caused by fading, shadowing, and by finally converting to dBm 
\begin{equation}
\text{RSSI}_l = 30+ 10\log_{10} {P_l} + X  \quad (\text{dBm})
\label{eq:rssi}
\end{equation}

%\Blue{(Gabriele, could you please edit the above and insert a short formulation and description about the RIS end-to-end model works, some equations, and how the matlab code  path-loss is used to generate Figure \ref{fig:map} ?. 
%E.g. similar to equation (2) in https://arxiv.org/pdf/2008.02459.pdf ) and from your paper https://arxiv.org/pdf/2009.02694.pdf 
%\\
%aim for 1-2 paragraphs please.}

\begin{figure}[t]
\centering
\includegraphics[ width=0.8\linewidth]{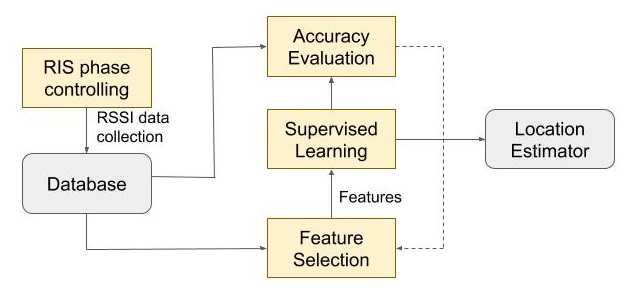}
\caption{Illustration of the proposed algorithm} \label{fig:framework}
\end{figure}

%%%%%
\squeezeup
\section{Heuristic State Selection}
\label{sec:hss}

Heuristically, we could argue that we can select $M< S$ RIS states such that the localization accuracy of the MU at any one of the $L$ sampled locations is on average improved, and that one way to do so is to select them by maximising differentiability.
%Intuitively, a set of RIS masks which can make the mutual information on spatial location largest is a good candidate. 
For example, if $M=2$ and $S=4$ corresponding to the RSSI spatial distributions shown in Figure \ref{fig:map}, then the RIS state corresponding to Figures \ref{fig:map}a) and b), should not be chosen together because they are too similar. 
On the other hand, one of them can probably be chosen with the ones corresponding to Figures \ref{fig:map} c) or d). 
To automate and remove bias from such decisions, a mathematical formulation of similarity and differentiability is needed.
To that end, we can write down a simple method for choosing the most differentiable RIS subset
\es{
\vect{M}^* =\argmax_{\vect{M}\subset \vect{S}} \frac{1}{L}\sum_{l=1}^{L} \Delta\big( \hat{\vect{R}_l} \big) 
\label{heuristic}
}
where we have defined $\vect{S}$ as the set of all $S$ possible RIS states, and $\vect{M}\subset \vect{S}$ as a candidate subset of configurations, with  $S=|\vect{S}|$ and $ M=|\vect{M}|$.
In equation \eqref{heuristic}, the $\Delta( \hat{\vect{R}_l})$ function returns a difference (dissimilarity) metric between a set of $M$ RSSI values at location $l$.
The simplest difference metric we can use is the Euclidean norm
\es{
\Delta\big( \hat{\vect{R}_l} \big) = \frac{2}{M(M-1)}\sum_{m\not=n} \| \hat{R}_m - \hat{R}_n \|^2
\label{diff}
}
thus giving us a heuristic state selection (HSS) method for the RIS.
We note that the search space of equation \eqref{heuristic} is quite large.
In fact, there are ${}^S C_M = \frac{S!}{M!(S-M)!}$ possible RIS configurations. 
While there are many computationally efficient ways to reduce the HSS search space (e.g., Genetic Algorithms), the dominant issue that we can anticipate with HSS is actually that this approach suffers from outlier bias. 
If just one out of the $L$ locations has a very high difference metric, then the whole radio map ranks highly.
As we will see later on, this leads to sub-optimal localization accuracy.

\squeezeup
\section{Machine Learning Approach}
\label{sec:proposed}

Instead of heuristically choosing the RIS configuration states, we propose here to leverage off-the-shelf ML tools to train and thus inform this selection process through a data-driven training phase.
The proposed machine learning algorithm is illustrated in Figure \ref{fig:framework}.
To map the problem at hand into the realm of ML we consider each RIS configuration as a feature and thus aim to select an optimal set of $M$ features out of a superset of $S$ possible RIS configurations. 
Selecting features carelessly might result in the feature set containing uninformative, irrelevant, and mutual redundant features information. 
%Thus, we aim to identify a subset of $M$ features where for any pair within the selected feature set, the feature–feature mutual information is minimum and feature–spatial mutual information is maximum, inspired by the original works of Peng et al. \cite{peng2005feature}.
%Thus, the feature selection (FS) technique that we apply is a type of optimization problem used to achieve a new subset of informative features. 

Since the set of all possible RIS configurations $\vect{S}$ is often too large of a set, we first try to approximate it by constructing a smaller set $\tilde{\vect{S}}$ with $\tilde{S}\ll S$ chosen at random.
For instance, $\tilde{\vect{S}}$ can be created by assigning load impedance values to the RIS dipoles, e.g., chosen from different finite support random distributions or that follow some specific pattern.

Next, we aim to identify a subset of $M \ll \tilde{S}$ features that leverages the effectiveness of the fingerprinting algorithm. 
Among FS approaches, a wrapper approach enhances accuracy because the optimal feature subset is compatible with the specific biases and heuristics of the learning algorithms \cite{yang1998feature}.
In the FS training phase we generate noisy RSSI radio maps for each of the $\tilde{S}$ features using the tools described in section \ref{sec:radiomap} (see equation \eqref{eq:rssi}).
Each radio map consists of $L\times \tilde{S}$ data points.
Finally, we divide the database of radiomaps into two subsets: a training subset, and a  validation subset. 
The easiest way to do this is to train FS of RIS configurations on $L_t$ locations and then validate on $L_v$, such that $L_t+L_v=L$.

In the FS phase, a supervised learning approach is applied to localize the MU on the $L_t$ candidate test locations, i.e., matching a newly generated noisy RSSI value to the training database set, while only using a subset of $M$ RIS configurations.
To avoid testing all ${}^{\tilde{S}} C_M$ possible combinations of the RIS (and corresponding radio maps) we employ a Genetic Algorithm \cite{yang1998feature} to guide the selection process with a fitness objective function that minimizes localization error.
%The supervised learning algorithm can be $k$ Nearest Neighbor ($k$-NN), NN, RF.
Thus, iterating the feature search process several times we end up with a set of $M$ RIS configurations chosen from $\tilde{\vect{S}}$ that tend to return the best localization accuracy performance. 

While our proposed ML approach is described and later tested using computer generated radio maps, the same ML-FS approach can be applied with real radio map measurements thus making our algorithm practical and easy to use.

\squeezeup
\section{Performance Evaluation and Simulations}
\label{sec:simulation}

We perform numerical simulations to evaluate the proposed localization techniques. 
We assume an indoor space of $20 \times 20$  $\text{m}^2$ where the AP is just outside the top left corner of the room, and the RIS comprising of $N=16$ equally spaced dipoles on a 4x4 grid located at the middle of the bottom wall (see Figure \ref{fig:system}). 
Both AP and MU are equipped with SISO omnidirectional antennas. 
In our simulations, the AP emits signals at frequency of 2.4 GHz, with a transmission power of 0.1 Watt.
We also suppose that there is a VLOS between AP-RIS and RIS-MU, but no direct LOS between AP-MU, which is reasonable if the AP is in a different room or outdoors, while the MU is attempting localization indoors.
Finally, a zero-mean Gaussian noise is added to each noise-free RSSI value with standard deviation of 3 dBm, namely $X \sim \mathcal{N}(0, 3^2)$ in \eqref{eq:rssi}.

\begin{figure}[t]
\centering
\includegraphics[ width=0.7\linewidth]{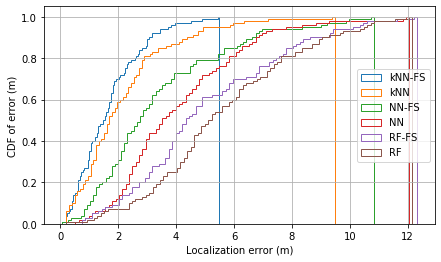}
\caption{CDF of the localization error using different supervised learning methods with and without using FS.} 
\label{fig:cdf}
\end{figure}

To compare HSS and ML-FS, we generated $\tilde{S}=50$ different RIS configurations and simulate the corresponding radio maps at $L=100$ (sparse 2x2 grid) and $L=400$ (dense 1x1m grid) locations.
The $\tilde{\vect{S}}$ set is chosen to include: 10 RIS configurations where the dipoles are set to the same value of the impedance thus emulating planar reflection, 10 configurations where the dipoles have a uniformly increasing impedance value thus emulating beamsteering, and 30 configurations in which the impedance values are randomly chosen thus emulating random (diffuse) scattering. 
This process allows us to create a diverse RIS configuration set and corresponding RSSI radio maps.
We then choose the optimal set $\vect{M}^*$ following the HSS approach \eqref{heuristic}
and the ML-FS approach on $L_t=L/10$ randomly chosen grid locations.

%\Blue{(Ly-san, here is the type of simulations and corresponding figures that I think we need to show that this works: (1) update figure 4 such that it compares the error cdf of (FS) vs (HSS) vs (random FS) vs (no RIS) using L=400, M=15. I expect that these are of decreasing localization performance, but that FS and HSS are close together thus proving our heuristics. (2) update figure 5a such that it compares (FS) vs (HSS) vs (random FS) vs (no RIS) as a function of M between 5 and 25 using L=100 (2x2). (3) update figure 5b such that it compares (FS) vs (HSS) vs (random FS) vs (no RIS) as a function of M between 5 and 25 using L=400 (1x1) (4) create a nuew figure 6 comprising of a heatmap (or 3D plot) of the mean localization error for (FS) as a function of N and D, using M=15 and L=400. This would provide useful engineering insights about what kind of RIS one should use: a large but cheaper one (N-big \& D-small), or a small and high quality one (N-small \& D-big)?
%Ly: I think we should not put too much results in this letter due to the limitation of the length. Besides, the computation time for doing (4) is to much, probably more than 1 week. Besides, even if experiment results show accuracy is enhanced using, say (N-big & D-small), we cannot conclude that (N-big & D-small) is good for any case, since it depends on many other factors.}

We perform two experiments. 
In the first experiment, we investigate three well-known supervised learning algorithms for localization using the FS-ML approach. 
Namely, we set $M=15$ and $\tilde{S}=50$, while $N=16$, $D= 200$, and $L=400$, and compute the cumulative distribution function (CDF) of the localization error under weighted $k$ Nearest Neighbors ($k$-NN), Neural Network (NN), and Random Forest (RF) \cite{han2011data}. 
For $k$-NN we set $k=5$ and the weight is set to the Euclidean distance. 
For NN,  we set the number and size of hidden layers to 1 and 100,  respectively, with an activation function set to the rectified linear unit function. 
For RF, the number of trees in the forest is set to 100, the function that measures the quality of a split is set to the Gini impurity, and nodes are expanded until all leaves are pure \cite{han2011data}.
%Definition of RF, if needed
%A random forest is a meta estimator that fits a number of decision tree classifiers on various sub-samples of the dataset and uses averaging to improve the predictive accuracy and control over-fitting. 

The results of the first experiment are shown in Figure \ref{fig:cdf} comparing localization  errors between the three matching algorithms ($k$-NN, NN, RF) with and without FS.
It is observed that $k$-NN performs the best out of the other three algorithms, and that FS always has a positive enhancement effect.

\begin{figure}[t]
\centering
\includegraphics[ width=0.6\linewidth]{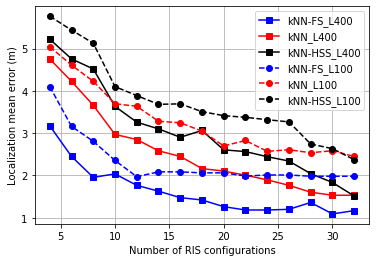}
\caption{Mean localization error vs. number of RIS configurations $M$. 
Blue squares, red squares, and black squares indicate mean localization error using 400 reference points, i.e. resolution of $1 \times 1 \text{m}^2$. Blue circles, red circles, and black circles indicate mean localization error using 100 reference points, i.e. resolution of $2 \times 2 \text{m}^2$. Blues squares/circles indicate the method that uses Feature Selection. Black squares/circles indicate the method that uses Heuristic State Selection.
Red squares/circles indicate the $5$-NN method using randomly chosen RIS configurations}
\label{fig:error}
\end{figure}

In the second experiment, we investigate the $k$-NN localization accuracy between FS and HSS as a function of $M$ using $L=100$ and $L=400$. 
We also plot the case where $M$ configurations a re-chosen at random from the simulated set of $\tilde{S}$.
The results for the second experiment are shown in Figure \ref{fig:error}.
As expected, we observe that all algorithms converge towards the grid size of 1 and 2 meters with increasing $M$.
Also, we observe that ML-FS performs the best, while the HSS is the worst, i.e., randomly choosing RIS configurations is actually better than our heuristic.
We ascribe this to a badly designed heuristic which suffers from outlier bias that tend to maximize the distance metric \eqref{diff}. 
Finally, we observe in Figure \ref{fig:error} that one can trade-off radio map resolution $L$ by applying FS, or by using a larger $M$.
For example, a mean localization error of 2 m can be achieved by having $L=400$ (dense) radio map grid points with $M=22$ random RIS configurations, or by using $L=100$ (sparse) grid points with ML-FS and $M=12$, thus saving both time and overall complexity.

%%%%
\squeezeup
\section{Conclusions and Discussion}
\label{sec:conclusion}
%summary

We have proposed and evaluated a novel and practical machine learning method for wireless fingerprinting localization in RIS-assisted environments.
The individual components of the proposed method such as $k$-NN localization and Genetic algorithms are off-the-shelf, have been validated previously in various settings, and are supported by standard machine learning software tools. 
What has enabled us to assemble and apply these components in an innovative way here, is the smart reconfigurable radio environment that is unlocked by RISs, thus replacing the need for having multiple APs and large numbers of fingerprint grid sample points while achieving great localization results.

This initial investigation suggests that RIS and smart radio environments can be exploited to attain sub-meter localization accuracy. 
To that end, future works should apply and further modify machine learning methods in more demanding and therefore optimizable settings, e.g., mixed environments with both LOS and NLOS, higher operational radio frequencies, more RIS elements ($N\gg 1$), and also using multiple RISs.

%%%%%
%\squeezeup
%\section*{Acknowledgments}
%This work is supported by 

%%%%
\squeezeup
\bibliographystyle{IEEEtran}
\bibliography{mybib}

\end{document}